\documentclass[twocolumn,aps,eqsecnum,graphicx]{revtex4}
\usepackage{graphics}
\newcommand{\be}{\begin{equation}}
\newcommand{\ee}{\end{equation}}

\newcommand{\bea}{\begin{eqnarray}}
\newcommand{\eea}{\end{eqnarray}}
\newcommand{\bd}{\begin{displaymath}}
\newcommand{\ed}{\end{displaymath}}
\newcommand{\bi}{\begin{itemize}}
\newcommand{\ei}{\end{itemize}}
\newcommand{\bc}{\begin{center}}
\newcommand{\ec}{\end{center}}
\newcommand{\bfl}{\begin{flushleft}}
\newcommand{\efl}{\end{flushleft}}
\newcommand{\bfr}{\begin{flushright}}
\newcommand{\efr}{\end{flushright}}
\newcommand{\f}{\frac}

\def\br{{\bf r}}\def\bs{{\vec \sigma}}
\def\bk{{\bf k}}  \def\bp{{\bf p}}
\def\bK{{\bf K}}\def\bR{{\bf R}}
\def\da{\downarrow} \def\ua{\uparrow} 

\def\6{\partial}  \def\b{\beta}
 \def\d{\delta} \def\ve{\varepsilon}

\def\ss{\sigma} 

\def\o{\omega}  \def\D{\Delta}

\def\={\!\!\!&=&\!\!\!}
\def\+{\!\!\!&&\!\!\!+~}
\def\-{\!\!\!&&\!\!\!-~}

\begin{document}

\author{Ionel \c{T}ifrea\cite{adresa}}
\affiliation{Department of Physics and Astronomy, University of
Iowa, Iowa City 52242, USA}
\author{Michael E. Flatt\'{e}}
\affiliation{Department of Physics and Astronomy, University of
Iowa, Iowa City 52242, USA}
\date{\today}
\title{Nuclear spin dynamics in parabolic quantum wells}

\begin{abstract}
We present a detailed analytical and numerical analysis of the
nuclear spin dynamics in parabolic quantum wells. The shallow
potential of parabolic quantum wells permits substantial
modification of the electronic wave function in small electric
fields.  The nuclear spin relaxation via the hyperfine interaction
depends on the electronic local density of states, therefore the
local nuclear relaxation time depends sensitively on the electric
field. For an inhomogeneous nuclear magnetization, such as
generated by dynamic nuclear polarization, the total nuclear
magnetization dynamics can similarly be altered. We examine this
effect quantitatively and the effect of temperature, field, well
thickness, and nuclear spin diffusion.
\end{abstract} \pacs{}\maketitle

\section{Introduction}

The long electronic spin coherence times in GaAs structures
\cite{om1} may assist the practical realization of spin based
electronics, including quantum information processing.\cite{book}
Electronic spin interactions with phonons, nuclear spins, other
electron spins and impurities will limit the value of the
electronic spin relaxation and coherence times. Due to much longer
coherence times nuclear spins can also be useful elements of
spin-based devices. Compared to electronic spin relaxation times,
of the order of 100 ns, nuclear spin relaxation times in the
absence of carriers are roughly 10 min in GaAs
structures.\cite{berg,barret,smet} Yet the local electronic
density of states near a nuclear spin can be increased to the
point where the hyperfine interaction  dominates nuclear spin
relaxation and the relaxation times are shorter than 1 s.

In this work we investigate the nuclear spin relaxation time in
 a parabolic quantum well (PQW). In this low
dimensional structure electrons are trapped in a parabolic
potential that restricts their motion along one direction. Such a
profile for the confining potential can be obtained by gradually
varying the Al concentration, $x$, of Al$_x$Ga$_{1-x}$As layers
between two barriers.\cite{parQW}   In the presence of an electric
field the minimum of the confining potential is displaced with
respect to the center of the PQW, whereas the shape of the
confining potential and implicitly of the envelope wave function
are largely conserved. This particular property of the PQW makes
it suitable for the practical realization of electronic devices
based on the electrical manipulation of the electronic or nuclear
spins. Recent experiments\cite{salis2} demonstrated that
considerable electrical control of {\it electronic} spin dynamics
($g$-factors)  can be achieved in shallow potential parabolic
quantum wells.

The standard technique used  to extract nuclear spin relaxation
times is nuclear magnetic resonance (NMR) spectroscopy. For  low
dimensional systems such as PQW's the signal in a standard NMR
experiment from the number of available nuclei across the sample
is too small to measure. The acquired polarization of nuclei in
such samples under a magnetic field of 10T is about
10$^{-4}-10^{-5}$ per nucleus, insufficient to generate signal
from a sample with 10$^{12}-10^{15}$ nuclei.\cite{barret} Higher
levels of nuclear polarization are required for a detectible NMR
signal. An alternative to standard NMR spectroscopy is optically
pumped NMR spectroscopy. In this approach non-equilibrium electron
spin polarizations are generated through exciting electronic
interband transitions. These spin-polarized electrons then
polarize the nuclei during their
relaxation.\cite{barret,NMR2,harley,salis1} This process is known
as dynamical nuclear polarization (DNP).\cite{abragam} DNP has
recently been achieved electrically in ferromagnetic-semiconductor
heterostructures.\cite{strand}

The nuclear spin coherence and relaxation dynamics will also be
influenced by the fact that all nuclei in GaAs carry spins, so
nuclear diffusion cannot be ignored.\cite{paget,harley} In the DNP
process the initial spatial distribution of the nuclear
polarization will follow the nuclear-electron mutual spin-flip
rate, proportional to the square of the electronic local density
of states, for the hyperfine interaction is a point contact
interaction. In lightly doped bulk semiconductors the initial
polarization of the nuclei is realized around impurities, whereas
in PQW's the confining potential creates a position-dependent
electron distribution, which will determine also the initial
polarization of the nuclei through the sample. Under continual
optical pumping the diffusion process will lead, eventually, to
the uniform polarization of the nuclei in the whole sample. The
value of the diffusion constant in GaAs systems was estimated
previously by Paget\cite{paget} for bulk samples and by Malinowski
and Harley\cite{harley} for square quantum well structures to be
of the order of 10$^{-14} - 10^{-13}$cm$^2$/s.

Once polarized, the nuclear spin dynamics can be controlled
directly with magnetic fields, or indirectly with electric fields.
A method for electric field control of the nuclear spin relaxation
time in quantum well (QW) structures was first reported by Smet
{\em et al.}\cite{smet}, and is based on electrically tuning the
electron density in a QW across a Quantum Hall ferromagnet
transition. In the low temperature regime such control is possible
due to electric field induced modifications to the spectrum of
collective excitations.

In this work we analyze two different situations in which the
nuclear spin relaxation time can be controlled with an external
electric field. We treat the two-dimensional electron gas at much
higher temperatures, outside of the Quantum Hall regime. The DNP
process remains very efficient up to 30K, and we predict a
sensitive dependence of the nuclear spin relaxation time on the
electric field, originating from the manipulation of the
electronic local density of states across the PQW.  This
manipulation of the electronic local density of states produces a
position-dependent nuclear spin relaxation rate. The position
dependence can be observed in a sample with a single
$\delta$-doped layer of an NMR active material inserted at a
certain position inside the PQW.  The inserted layer needs to be
made of different nuclei than the one present in the host PQW to
ensure that its observed resonant frequency is not the same as
that of the host PQW nuclei. Even in a sample without a specific
$\delta$-doped layer substantial control of the nuclear relaxation
rate is possible. For uniformly distributed nuclei in the PQW, the
appropriately averaged nuclear relaxation time depends on the
occupancy factor of the electronic conduction subbands, decreasing
stepwise as the electron density increases.

We report both analytical and numerical results for the electric
field dependent nuclear spin relaxation time. The special
structure of the confining potential in the PQW allows exact
analytical solutions for the position dependent electronic density
across the structure. More accurate results can be obtained from
numerical calculations of the electronic local density of states
in the PQW based on a {\bf k$\cdot$p} code, however we find that
the analytic results are respectably accurate for the structures
we considered. We find that the electron and nuclear spin
relaxation times can be tuned over a range of at least an order of
magnitude using an external electric field. To identify the most
suitable PQW structure for the nuclear spin relaxation time
manipulation, we investigate two different possible samples, with
different confining potentials.

The paper is organized as follows. The second section is dedicated
to the general presentation of the calculation of the nuclear
relaxation time as a result of the hyperfine interaction. The
third section presents our analytical and numerical analysis of the
PQW. Finally we summarize our results in the last section.

\section{The general formalism for the electronic and nuclear relaxation times}

Our analysis of the electronic and nuclear spin relaxation times in
low dimensional systems follows the previous calculation by
Overhauser\cite{over2} for the case of three dimensional (3D) bulk
metals. We only consider effects due to the hyperfine interaction
which couple the electronic and nuclear spins. The interaction
Hamiltonian, known also as the Fermi contact term, can be written
as
\be\label{intham}
H=\sum_n H(\br_n) = \sum_n \f{8\pi}{3}\;\b_e\b_n\left({\vec \ss_n }\cdot{\vec
\ss_e}\right)\;\d(\br-\br_n)\;,
\ee
where $\b_n$ and $\b_e$ are the nuclear and electron spin magnetic
moments,  ${\vec \ss_n}$ and ${\vec \ss_e}$ the Pauli spin
operators for the nucleus and electron, and $\d(x)$ is the usual
delta function. The argument of the delta function in Eq.
(\ref{intham}), $\br-\br_n$, represents the relative distance
between the nuclear and electronic spins, and shows the local
character of the hyperfine interaction. The effects of the Fermi
contact term are better understood if we express the product of
the two Pauli spin operators, $(\bs_n\cdot\bs_e)$, in terms of the
creation and annihilation operators, $\ss^\pm_{n(e)}$, as
\be\label{pauliop}
\bs_n\cdot\bs_e=\ss_n^z\ss_e^z+2\left(\ss_n^+\ss_e^-+\ss_n^-\ss_e^+\right)\;,
\ee
an expression which clearly identifies the spin-flip processes
involving both the electron and nuclear spin. The second term in
Eq. (\ref{pauliop}) flips a nuclear spin from down to up in a
process which implies also an electron spin flip from up to down.
The last term in Eq. (\ref{pauliop}) describes the reverse
process.

To understand the spin relaxation process one has to consider the
effects of a magnetic field $\cal H$ acting on the particle's
spin. The final result of the interaction between the spin and the
magnetic field will be an orientation of the spin, parallel or
antiparallel, relative to the applied field. However, such a
process does not occur instantaneously, and its time dependence
defines the characteristic spin relaxation time. The  dependence
on different parameters of the relaxation time originates from the
particular mechanism which produces the spin alignment. In our
case this mechanism will be the hyperfine interaction. As the
Fermi contact term involves both the electron and the nuclear
spins we expect to obtain a single equation which will determine
both the electron and nuclear spin relaxation times.

The approximations we introduce are natural ones for
nanostructures. We consider the electronic system to be in spin
equilibrium with itself. Thus even though the electronic local
density of states may vary from position to position, the
electronic spin polarization is the same everywhere. Situations
where the electronic spin system has position-dependent spin
polarization can be considered by combining the types of
expressions found here with electron spin diffusion equations. For
our purposes here we assume that at any specific time we have
$N_+$ of the electrons with their spins oriented along the applied
field (spin up) and $N_-$ with their spins oriented antiparallel
to the applied field (spin down).

In contrast to the electronic system we permit the nuclear system
to develop spatially-dependent polarization. For simplicity we
will derive the expressions for electronic and nuclear spin
relaxation times considering only nuclei with total spin 1/2; the
generalization for other nuclear spin values is straightforward.
Thus for a given nucleus the probability of being spin up is
$M_+(\br_n)$ and of spin down is $M_-(\br_n)$.    One might be
concerned if these local polarizations are well defined -- it is
known that in some circumstances non-Markovian effects can play a
role, and we ignore such effects here. This approximation can be
interpreted in some circumstances as averaging over sufficiently
long times, and in other circumstances of coarse-graining the
nuclear system over a volume large compared to the internuclear
distance, but small compared to relevant electronic length scales.
This coarse-graining permits us to replace the orientation of a
given nucleus in these equations with a statistical average.

The time dependence of the relaxation process will be related to
the time evolution of both the total electronic magnetization and
the position-dependent nuclear magnetization, which can be
expressed, in terms of the differences, $D=N_+-N_-$, and
$\D(\br_n)=M_+(\br_n)-M_-(\br_n)$. If we denote by $W_{+-}(\br_n)$
the total numbers of electron spins which flip from down to up per
second, and by $W_{-+}(\br_n)$ the total number of electron spins
which flip from up to down per second, each from interaction with
nucleus $n$, we can write the change in $D$ as
\be\label{Dprob}
\f{dD}{dt}=2\sum_n[W_{+-}(\br_n)-W_{-+}(\br_n)]\;,
\ee
and in $\D(\br_n)$ as
\be\label{dDprob}
\f{d\D(\br_n)}{dt} = 2[W_{-+}(\br_n)-W_{+-}(\br_n)]
\ee
The evaluation of $W_{+-}(\br_n)$ and $W_{-+}(\br_n)$ will be done based on
Fermi's golden rule by treating the hyperfine interaction as a
time dependent perturbation.

Consider an electron transition between two states $\bk_\da$ and
$\bk_\ua'$ in which the electron spin is flipped from down to up.
According to Fermi's golden rule, the electronic transition
probability from state $\bk$ to state $\bk'$ induced by nucleus
$n$ is given by
\be\label{FGR}
W^{\da\ua}_{\bk\bk'}(\br_n)=\f{2\pi}{\hbar}\;\left|H_{\bk\bk'}(\br_n)\right|^2\;
N_e(\bk_\ua')\;\d(E_i-E_f)\;,
\ee
where
$H_{\bk\bk'}(\br_n)=\left<\psi_f(\bk_\ua')\left|H(\br_n)\right|\psi_i(\bk_\da)\right>$
represents the matrix elements of the hyperfine interaction
Hamiltonian for nucleus $n$, $N_e(\bk_\ua')$ is the electronic
density of states for the final state, and $E_i$ and $E_f$ are the
energies corresponding to the initial and final state.
Accordingly, the total number of spin flips per second induced by
nucleus $n$, $W_{+-}(\br_n)$, is
\bea\label{probdownup}
&&W_{+-}(\br_n)=\nonumber\\
&&\sum_{\bk} W^{\da\ua}_{\bk\bk'}(\br_n) M_+(\br_n)
f_{FD}(\bk_\da, E_i) \left[1-f_{FD}(\bk_\ua', E_f
)\right]\;,\nonumber\\
\eea
where $f_{FD}(\bp_\ss,E)$ is the usual Fermi-Dirac distribution
function for an electron with momentum $\bp$ and spin $\ss$. The
energy conservation expression
\be \label{encons}
\f{\bk_\ua'^2}{2m}-\b_e{\cal H}-\b_n{\cal
H}=\f{\bk_\da^2}{2m}+\b_e{\cal H}+\b_n{\cal H}\;,
\ee
allows us to eliminate $\bk_\ua'$ and estimate $W_{+-}(\br_n)$ using Eq.
(\ref{probdownup}) by replacing the sum over the momenta with an
integration over the energy using the density of states. The
integration over energies can be performed simply in the case of
small magnetic fields, $\b_{e(n)}{\cal H}\ll k_BT$ ($T$ is the
temperature), for which the Fermi-Dirac function can be expanded
in a Taylor series. $W_{-+}(\br_n)$ is calculated in a similar way using
$M_-(\br_n)$ instead of $M_+(\br_n)$ and changing $\cal H$ to $-\cal H$.

After some algebra we find that the time dependence of the
electron magnetization is described by the following equation
\bea\label{Dfinal}
&&\f{dD}{dt}=\sum_n\f{512\pi^3\b_n^2\b_e^2\tilde{N}_e
|\psi(\br_n)|^4}{9\hbar V}(D_0-D)\nonumber\\&&+
\sum_n\f{1024\pi^3\b_n^2\b_e^2\tilde{N}^2_e k_BT
|\psi(\br_n)|^4}{9\hbar }(\D_0(\br_n)-\D(\br_n))\;,\nonumber\\
\eea
where $\tilde{N}_e=\int d\ve N_e(\ve) f_{FD}'(\ve)$, $D_0$ and
$\D_0(\br_n)$ are the thermal equilibrium values for $D$ and
$\D(\br_n)$, respectively, and $\psi(\br_n)$ is the electronic
wave function at the $n$'th nuclear position. The first term on
the right hand side (rhs) of Eq. (\ref{Dfinal}) determines the
electron spin relaxation time, whereas the second one the
position-dependent nuclear spin relaxation time. The
generalization to the case of arbitrary nuclear spin $I$ is
straightforward if we consider that the nuclear spin flips
probabilities $W_{-+}(\br_n)$ and $W_{+-}(\br_n)$ describe
transitions between adjacent nuclear spin levels.

In view of Eq. (\ref{Dfinal}) the definitions of the electronic
and nuclear spin relaxation times should be considered with some
caution, as an exponential decay for the electronic magnetization
is obtained only for the case when the nuclear population
approaches equilibrium, or is kept at a fixed nonequilibrium
value. For the latter case a new kinetic equilibrium value for the
electron polarization, $\tilde{D}_0$, will be reached, which is
determined by the value for $D$ that makes $dD/dt=0$ in Eq.
(\ref{Dfinal}) for the constant nonequilibrium nuclear
polarization $\D(\br_n)$. For calculating the nuclear spin
relaxation time we use the total spin angular momentum
conservation
\be\label{cons}
\f{dD}{dt} = \f{2I(I+1)(2I+1)}{3}\sum_n \f{d\D(\br_n)}{dt}\;,
\ee
assume the electron spin polarization is refreshed, and so remains
approximately a constant, and separate the resulting equation into
$n$ equations: one for each nucleus.

For the case of a two dimensional (2D) system, with a constant
density of states, $N_e(\ve)=m_e/2\pi$ ($m_e$ is the electron
mass), the electronic and nuclear relaxation times are
\be\label{te}
T^{-1}_{1e}=\f{1}{A}\sum_n\f{512\pi^2\;m_e\;\b_e^2\;\b_n^2\;
\left|\psi(\br_n)\right|^4}{9\hbar^3\;(2I+1)}
\ee
and
\be\label{tn}
T^{-1}_{1n}(\br_n)=\f{128\pi\;m_e\;\b_e^2\;\b_n^2\;\left|\psi(\br_n)\right|^4
k_BT}{3\hbar^5\;I(I+1)(2I+1)}\;,
\ee
where $A$ represents the area of the 2D system. Temperature
corrections to these times are of the order of $k_BT/E_F$, $E_F$
being the Fermi energy of the electronic system. As we can see
from Eq. (\ref{te}) the electronic spin relaxation time is
temperature independent (up to the first order of approximation)
suggesting that the relaxation mechanism due to the hyperfine
interaction can be the dominant one at low temperatures, as the
relaxation times corresponding to other mechanisms increase as the
temperature decreases.\cite{over2} According to Eq. (\ref{tn}) the
nuclear spin relaxation time follows the well known Korringa law,
which states that the ratio $1/(T_{1n}T)$ is temperature
independent.

Let us turn our attention to the quasi-two-dimensional case of a
QW. In such systems the growth-direction electronic motion is
restricted to discrete energy levels through confinement in a
potential well. Accordingly, the electron energy dispersion can be
written as $\ve(\bk)=\ve_i+\bk^2/2m_e$, where $\ve_i$ is the
minimum of the conduction subband $i$ and $\bk$ is the electron
momentum in the plane of the quantum well. If the separation
between the energy levels is large enough compared to the Fermi
energy, the electrons will be frozen in the first energy subband
leading to the practical realization of a 2D electronic system. In
this case the electronic and nuclear spin relaxation times are
given by Eqs. (\ref{te}) and (\ref{tn}). However, if the Fermi
energy exceeds the separation between the energy levels, we must
consider a multiply-occupied subband system. For the case of
multiple subband occupancy the density of states is
\be\label{DOSQW}
N(\ve)=\left\{\begin{array}{ll}m_e/2\pi, & \ve_1<E_F<\ve_2\\
2m_e/2\pi, & \ve_2<E_F<\ve_3\\ \cdots\\
lm_e/2\pi, & \ve_l<E_F<\ve_{l+1}\end{array}\right.\;,
\ee
where we consider that the $l$-th is the last subband of the
system occupied by electrons. Two additional complications arise
that prevent direct use of this larger density of states in
Eqs.~(\ref{te}-\ref{tn}). First, the envelope functions for the
different subbands will differ, and second, electron spin-flip
scattering can take place inside the same subband or between two
different subbands.

The electron and nuclear spin relaxation times for a QW can still
be written very simply, and physically, in terms of the electronic
local density of states,
\be\label{teQW}
T^{-1}_{1e}=\f{1}{V}\sum_{n}\f{1024\pi^3\b_e^2\b_n^2\int d\ve
A_e^2(\br_n,\ve) f_{FD}'(\ve)}{9\hbar (2I+1)\int d\br d\ve
A_e(\br,\ve) f_{FD}'(\ve)}
\ee
and
\be\label{tnQW}
T^{-1}_{1n}(\br_n)=\f{512\pi^3\b^2_e\b^2_nk_BT\int d\ve
A_e^2(\br_n,\ve) f_{FD}'(\ve)}{3\hbar I(I+1)(2I+1)}\;.
\ee
The electronic local density of states at the nuclear position
$\br_n$ is
\be
A_e(\br_n,\ve)=\sum_m |\psi_m(\br_n)|^2\d(\ve-E_m)\;,
\ee
where $m$ labels the state, $E_m$ its energy, and $\psi_m(\br_n)$
its wavefunction at the $n$'th nucleus.

In general for any QW structure the dispersion relation is
quasi-two-dimensional, therefore, the electronic wave function
will be written as a product between an envelope function,
$\phi(z)$, and a Bloch function, $u_{n\bK}(\br)$, such that
$\psi_{\bK,n}(\br_n)=\exp{[i\bK\cdot\bR]}\phi(z)u_{n\bK}(\br_n)$,
and
\be
A_e(\br_n,\ve)=\sum_{j\bk}|\psi_{j\bK}(\br_n)|^2\d(\ve-E_{j\bK})\;,
\ee
where $j$ is the subband index and $\bK$ the momentum. The
envelope functions for parabolic quantum wells will be evaluated
both analytically and numerically. For the numerical estimations
we will use a $\bk\cdot\bp$ calculation.\cite{thesis} For the QW's
stepwise density of states,
\bea\label{ae}
&&\int d\ve A_e^2(\br_n,\ve) f_{FD}'(\ve)\nonumber\\
&&=|u(\br_n)|^4\sum_{j,k}|\phi_j(\br_n)|^2|\phi_k(\br_n)|^2
\Theta(\ve_{max\{j,k\}}) \;,\nonumber\\
\eea
with $\Theta(\ve_j)$ a temperature dependent factor. Such factors
are important when the Fermi energy approaches an $\ve_i$, i.e.,
when the electrons start to occupy higher conduction subbands. The
energy integration over the step-like density of states in Eq.
(\ref{ae}) leads to temperature dependent factors of the following
form:
\be\label{tempfact}
\Theta(\ve_j)=\f{1}{\exp{\left[\f{\ve_j-E_F}{k_BT}\right]}+1}\;.
\ee
Eq. (\ref{tempfact}) clarifies the role of the temperature when
higher subbands start to be occupied, i.e., the Fermi energy is of
the order of $\ve_i$.

For $\bK\sim 0$ the value of the Bloch function for the conduction
band can be approximated as a constant. The approximate constant
value of the Bloch function at the $^{71}$Ga nucleus site,
$|u(\br_n)|^2=5.2\times10^{25}$ cm$^{-3}$, was extracted by
comparing our numerically calculated nuclear relaxation time with
the available experimental data for a GaAs/Al$_{0.1}$Ga$_{0.9}$As
QW.\cite{tifrea}

\section{The Parabolic Quantum Well}

In this section we will discuss the nuclear spin relaxation time
for the case of parabolic QW's. The special form of the confining
potential in these structures allows for exact analytical results
which permit an evaluation of the nuclear spin relaxation time.
The analytical results will be compared with numerical data
obtained from the $\bk\cdot\bp$ calculation.

An initial nuclear polarization obtained by DNP will be
inhomogeneous, and for short times will be proportional to
$T^{-1}_{1n}(z)$, so for one occupied subband the initial nuclear
magnetization
\be\label{maginit}
m(z,t=0)\propto |\phi(z)|^4.
\ee
Our calculation assumes that DNP is performed in the absence of an
applied electric field and for single subband occupancy. In this
case $\phi(z)$ corresponds to electrons situated in the first
subband of the conduction band. The value of the nuclear
magnetization will be influenced mainly by two processes, spin
relaxation and spin diffusion. We track the evolution of the
position dependent nuclear polarization in order to obtain the
time dependence of the total nuclear magnetization in the PQW.
Considering both spin diffusion and spin relaxation, the total
nuclear magnetization across the PQW can be evaluated as the
solution to the following equation:
\be\label{diffeq}
\f{d m(z,t)}{d t}=D\f{\6^2m(z,t)}{\6z^2}-\f{m(z,t)}{T_{1n}(z)}\;,
\ee
where $D$ represents the nuclear diffusion constant. The combined
effects of nuclear spin diffusion and relaxation indicate that the
longer-time dynamics of the magnetization is non-exponential in
PQW's.\cite{tifrea}   If we consider that initially the PQW is in
a state where only the first conduction band is filled with
electrons, during the DNP procedure nuclei situated close to the
sample's center become strongly polarized. The initial
polarization is smaller for nuclei farther away from the center of
the PQW where the electronic density is smaller. Even farther from
the PQW center, where the electronic wave function is essentially
zero, the nuclei are initially unpolarized. However, nuclear spin
diffusion can bring the polarization generated in the center of
the well out to the entire sample.

The position dependent {\em initial} nuclear spin relaxation time
can be extracted from Eq. (\ref{diffeq}) if the nuclear diffusion
term is absent ($D=0$). As a consequence the observed total
initial nuclear relaxation time is given by
\be\label{average}
\f{1}{T_{1n}}=\f{\int dz\;T^{-1}_{1n}(z)\;{\cal P}_n(z)}{\int
dz\;{\cal P}_n(z)}\;,
\ee
where ${\cal P}_n(z)$ is proportional to the initial
position-dependence of the nuclear polarization. For the case of
optical pumping, where the nuclear polarization is realized
through the hyperfine interaction, ${\cal P}_n(z)\propto
|\phi_1(z)|^4$. Our estimation of the nuclear spin relaxation
times will be done for temperatures where DNP in GaAs is efficient
($T\sim 30$K), however, we expect that the obtained results can be
easily extended to higher temperatures.

\subsection{Analytical Results}

In an ideal parabolic QW (IPQW) the form of the confining
potential is
\be\label{potQW}
V(z)=\f{1}{2}m_e\o_0^2z^2\;,
\ee
where $\o_0$ is the characteristic frequency. The electronic
envelope functions and the characteristic subband energies are
obtained from the time-independent Schrodinger equation
\be\label{Schrodingereq}
\left(-\f{\hbar^2}{2m_e}\f{d^2}{dz^2}+\f{1}{2}m_e\o_0^2z^2\right)\phi(z)=\ve_n\phi(z),\
\ee
as
\be\label{envfunct}
\phi_{n}(z)=N_{n}\exp{\left[-\f{m_e\o_0}{2\hbar}z^2\right]}
H_{n-1}\left[\left(\f{m_e\o_0}{\hbar}\right)^{1/2}z\right]
\ee
and
\be\label{subbanden}
\ve_n=\left(n-\f{1}{2}\right)\hbar\o_0\;,
\ee
where the subband index $n$ is a  natural number, $H_n(x)$ are the
Hermite polynomials, and $N_n$ is the normalization factor which
can be obtained analytically if the width of the QW is much bigger
than the PQW length scale $L\gg l_0$
($l_0=\sqrt{\hbar/(m_e\o_0)}$). According to Eq. (\ref{subbanden})
the energy distance between adjacent conduction subbands, $\D
E=\ve_n-\ve_{n-1}$, is a constant depending only on the confining
potential's characteristic frequency $\o_0$.

An additional external electric field, $\cal F$, will change the
Schrodinger equation by introducing a field dependent term of the
form
\be\label{potfield}
V(z)=e{\cal F}z\;,
\ee
whose effects can be easily understood if  absorbed into the
parabolic confining potential. The external electric field will
not change the form of the envelope functions, however their
positions in the QW are shifted by a distance proportional to the
strength of the field, $z_0=e{\cal F}/m_e\o_0^2$. The new wave
functions are
\bea\label{waveelectric}
\phi_{n}(z, {\cal
F})&=&N_{n}\exp{\left[-\f{m_e\o_0}{2\hbar}(z-z_0)^2\right]}\nonumber\\
&&\times
H_{n-1}\left[\left(\f{m_e\o_0}{\hbar}\right)^{1/2}(z-z_0)\right]\;.
\eea
The subband energy in the presence of an external electric field
will be lowered to a value
\be\label{subbandenfield}
\ve_n({\cal F})=\ve_n-\f{e^2{\cal F}^2}{2m_e\o_0^2}\;,
\ee
where the energy difference, $\D E=\ve_n-\ve_{n-1}$, is the same
as in the absence of the electric field. The electric field will
allow a direct control of the electron density profile across the
sample leading to the manipulation of the nuclear spin relaxation
times in PQW's.

In the following we will consider two different PQW
configurations, i.e, two structures with different characteristic
frequencies $\o_0$: the first structure will be referred to as
IPQW I  ($\o_0$= 2.3$\times$10$^{13}$ s$^{-1}$, $\D E$ = 15 meV)
and the second one  as IPQW II ($\o_0$= 7.6$\times$10$^{13}$
s$^{-1}$, $\D E$ = 50 meV). This choice of the PQW's will allow us
to discuss the relative influences of temperature, electric field,
and well structure on the nuclear spin relaxation times.

\begin{figure}[t]
\vspace{0.5cm} \centering
\scalebox{0.3}[0.4]{\includegraphics*{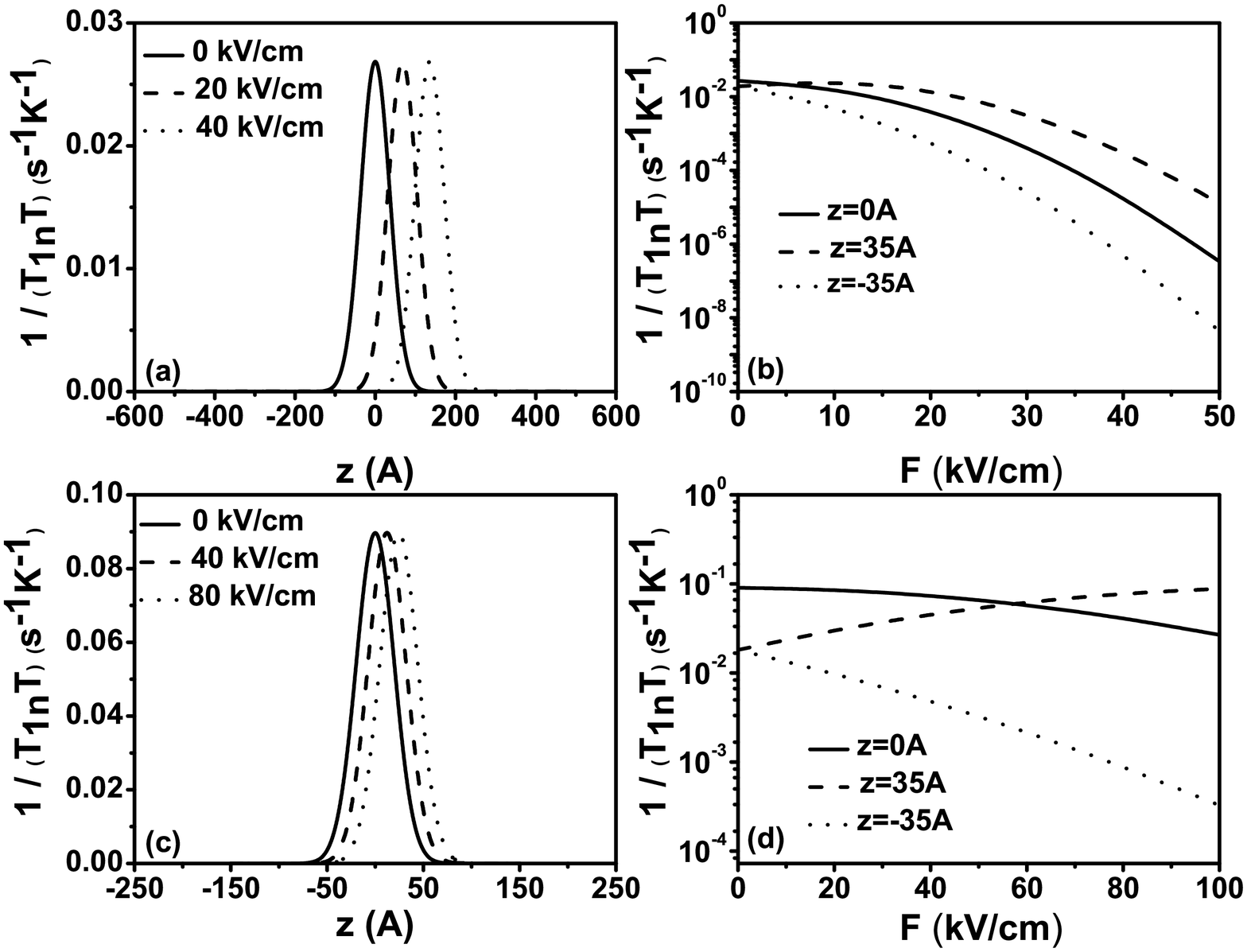}}
\caption{The position dependence of $1/(T_{1n}T)$ for different
values of the external electric field [(a)  IPQW I and (c)  IPQW
II]. The field dependence of the $1/(T_{1n}T)$ ratio for
$\d$-doped layers situated at different positions across the QW
[(b)  IPQW I and (d)  IPQW II].}
\label{fig2}
\end{figure}

In Fig.~\ref{fig2}(a) and \ref{fig2}(c) we present the position
dependence of $1/(T_{1n}T)$ as a function of the applied field for
the case of single subband occupancy. IPQW I is more suitable than
IPQW II for electric field control, as the rate of change with
field manifest in $1/(T_{1n}T)$ is much larger. The displacement
of the electronic wave function with an applied electric field is
larger for wells with smaller values of the characteristic
frequency, as $z_0\sim \o_0^{-2}$, and IPQW I has a smaller $\o_0$
than IPQW II. In Fig.~\ref{fig2}(b) and \ref{fig2}(d) we plot the
value of the $1/(T_{1n}T)$ ratio as a function of the applied
electric field for three different positions in the QW. These
positions corresponded to initial nuclear polarization via DNP of
the maximal value (center of the well) and half of the maximal
value. The tunability of the nuclear spin relaxation times at
these particular position is a few orders of magnitude, with
larger tunability for IPQW I.

\begin{figure}[t]
\vspace{0.5cm} \centering
\scalebox{0.3}[0.4]{\includegraphics*{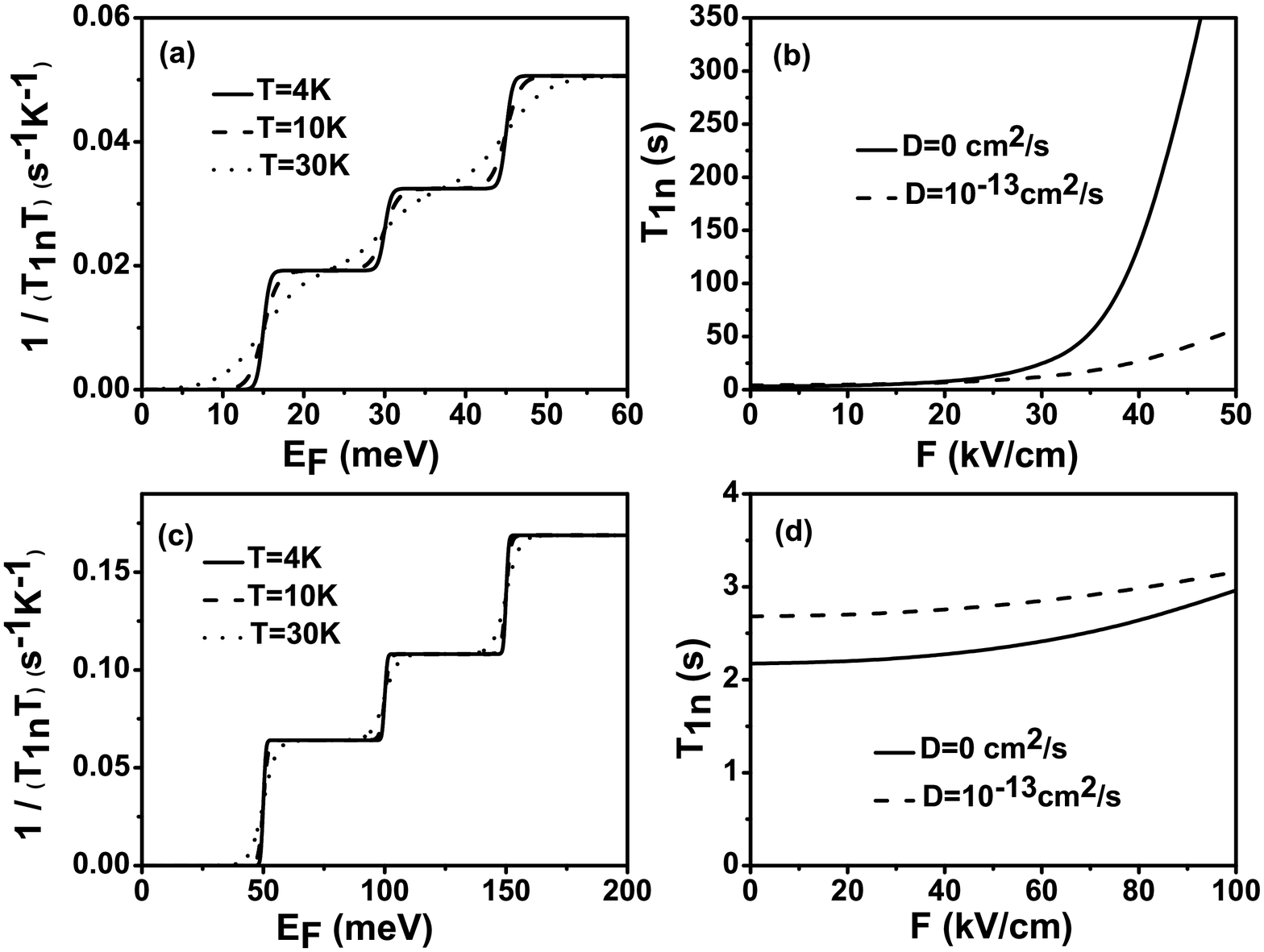}}
%
%
\caption{The \emph{initial} value of  $1/(T_{1n}T)$ as a function
of Fermi energy at several temperatures [(a) IPQW I and (c) IPQW
II]. The electric field dependence of the \emph{initial} nuclear
spin relaxation times for single subband occupancy, with and
without  nuclear spin diffusion [(b) IPQW I and (d) IPQW II].}
\label{fig1}
\end{figure}

This property of the PQW's will allow direct control of the
relaxation process in $\d$-doped layers of nuclei situated at
certain positions inside the QW. The only requirement  is that the
$\d$-doped layer is situated in a region where the DNP procedure
will produce a high enough initial polarization of the nuclei. The
nuclei inside these layers will not be affected by nuclear
diffusion, as this process requires similar nuclear species, or at
least nuclear species with close values of the resonance
frequencies.\cite{abragam}

In Fig.~\ref{fig1}(a) and \ref{fig1}(c) we present $1/(T_{1n}T)$
for  uniformly distributed nuclei as a function of the Fermi
energy (which depends on the electron density). Manipulation of
the electronic density in the QW can be accomplished with a gate.
The reference value for the energy is set at $-\D E$ below the
minimum of the first conduction band. $1/(T_{1n}T)$ depends
stepwise on the Fermi energy at very low temperatures, increasing
every time a new conduction electron subband becomes occupied.
However corrections to the Korringa law become important at higher
temperatures for Fermi energies near these steps. As we can see
from Fig.~\ref{fig1}(a) at T=30 K for IPQW I, temperature-induced
smearing will destroy the stepwise character of $1/(T_{1n}T)$. For
IPQW II due to a larger energetic separation between subbands the
stepwise behavior remains for higher temperatures.

Fig.~\ref{fig1}(b) and \ref{fig1}(d) present the {\em initial}
value of the nuclear spin relaxation time as a function of the
applied electric field. The relaxation time, both in the absence
of and presence of nuclear diffusion, is extracted as the first
derivative of the total nuclear magnetization in the QW. For a
nuclear spin diffusion constant close to the one extracted from
observations in bulk GaAs systems, the diffusion process will
diminish significantly the electric field tunability of the
nuclear spin relaxation time. Due to the larger distances the
electronic wave function is displaced in IPQW I, the tuning
calculated there is significantly more robust to nuclear spin
diffusion than IPQW II.

Thus we see that different PQW structures behave in a
complementary fashion in the presence of external electric fields.
If the electric field is used as an electron density gate,
controlling the subband occupancy in the system, the stepwise
dependence of $1/(T_{1n}T)$ on electron density can be better
observed in structures with a larger splitting of the energy
levels, for larger subband separations suppress the temperature
dependence of the steps. On the other hand, if  the electric field
is used to tune the nuclear spin relaxation time by controlling
the electronic local density of states (tilting the PQW),
structures with a smaller value of the characteristic frequency,
and implicitly with a smaller splitting of the energy subbands,
are more resilient to nuclear diffusion. When the field is used to
tilt the PQW, inserting a $\d$-doped layer of a different material
at a certain position across the QW will permit sensitive control
of the relaxation time over a few orders of magnitude (see Fig
\ref{fig2} (b) and (d)). The $\d$-doped layer should be made of a
different material, with a different NMR resonance frequency than
the host material.

\subsection{Numerical Results}

\begin{figure}[t]
\vspace{0.5cm} \centering
\scalebox{0.3}[0.4]{\includegraphics*{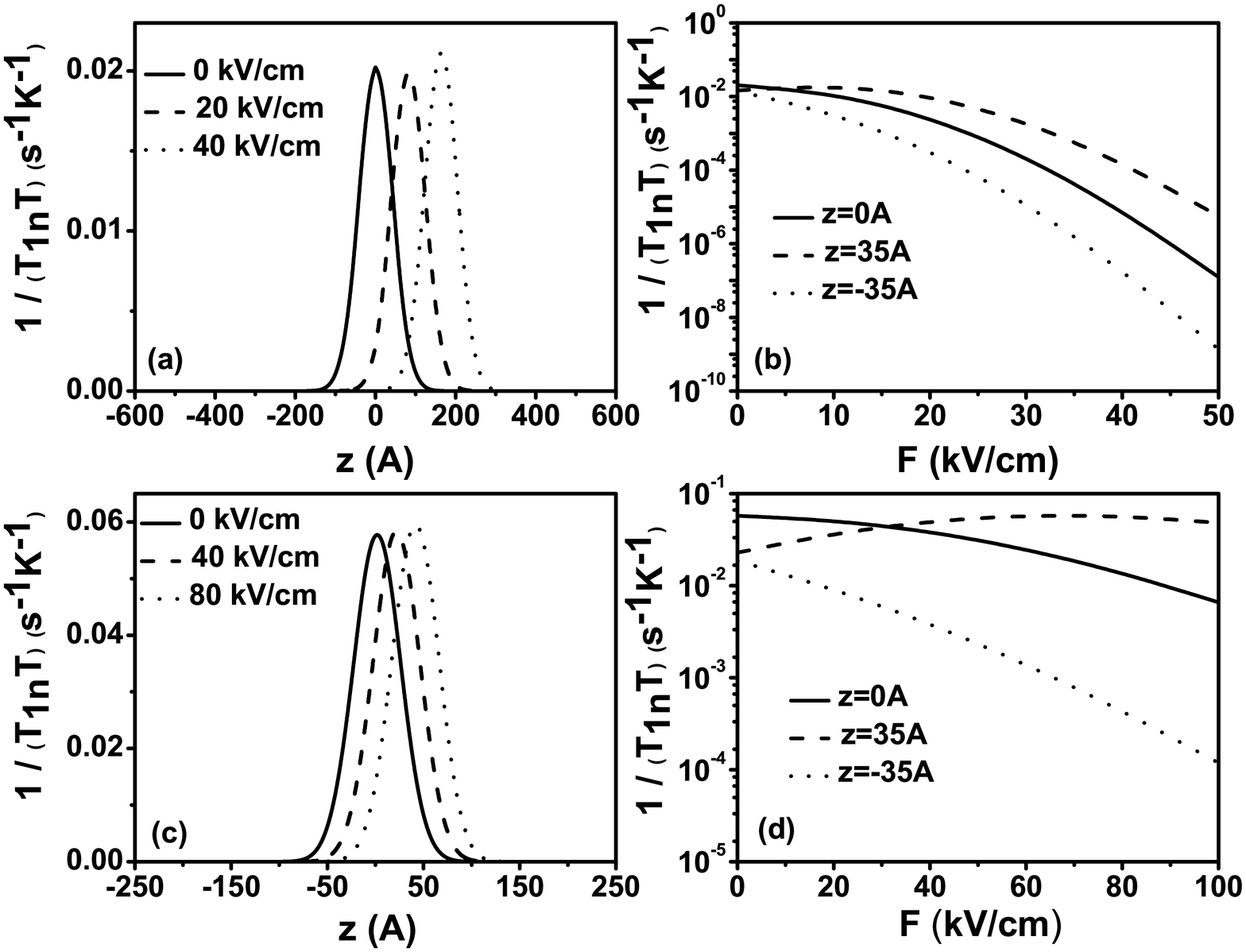}}
%
%
\caption{The position dependence of $1/(T_{1n}T)$ for different
values of the external electric field [(a) PQW I and (c) PQW II].
The field dependence of $1/(T_{1n}T)$ for $\d$-doped layers
situated at different position across the QW [(b) PQW I and (d)
PQW II].}
\label{fig4}
\end{figure}

In this section we will numerically evaluate nuclear spin
relaxation times for more realistic versions of the PQW's.  The
first structure (referred from now on as PQW I) is a 1000 \AA\
Al$_x$Ga$_{1-x}$As system sandwiched between two 100 \AA\
Al$_{0.4}$Ga$_{0.6}$As barriers.\cite{salis2} The value of the
aluminum concentration across the QW varies from 0.4 in the
barriers to 0 in the QW's center. The second structure (referred
from now on as PQW II) is a 300 \AA\ Al$_x$Ga$_{1-x}$As system
sandwiched between two 100 \AA\ Al$_{0.3}$Ga$_{0.7}$As
barriers.\cite{parQW} In this case the aluminum concentration
across the sample varies from 0.3 in the barriers to 0 in the
center of the QW. Our estimation of the energy difference between
the first and second conduction bands is $\D E=16.6$ meV for PQW I
and $\D E=46.1$ meV for PQW II, values which are close to the ones
considered in the two analytic examples. The evaluation of the
envelope functions in these particular structures is done based on
a 14-band multilayer ${\bf k} \cdot {\bf p}$ calculation.
Therefore additional effects will be evident in the ${\bf k} \cdot
{\bf p}$ calculations, including  a diminished value of the energy
difference between higher conduction subbands.

\begin{figure}[t]
\vspace{0.5cm} \centering
\scalebox{0.3}[0.4]{\includegraphics*{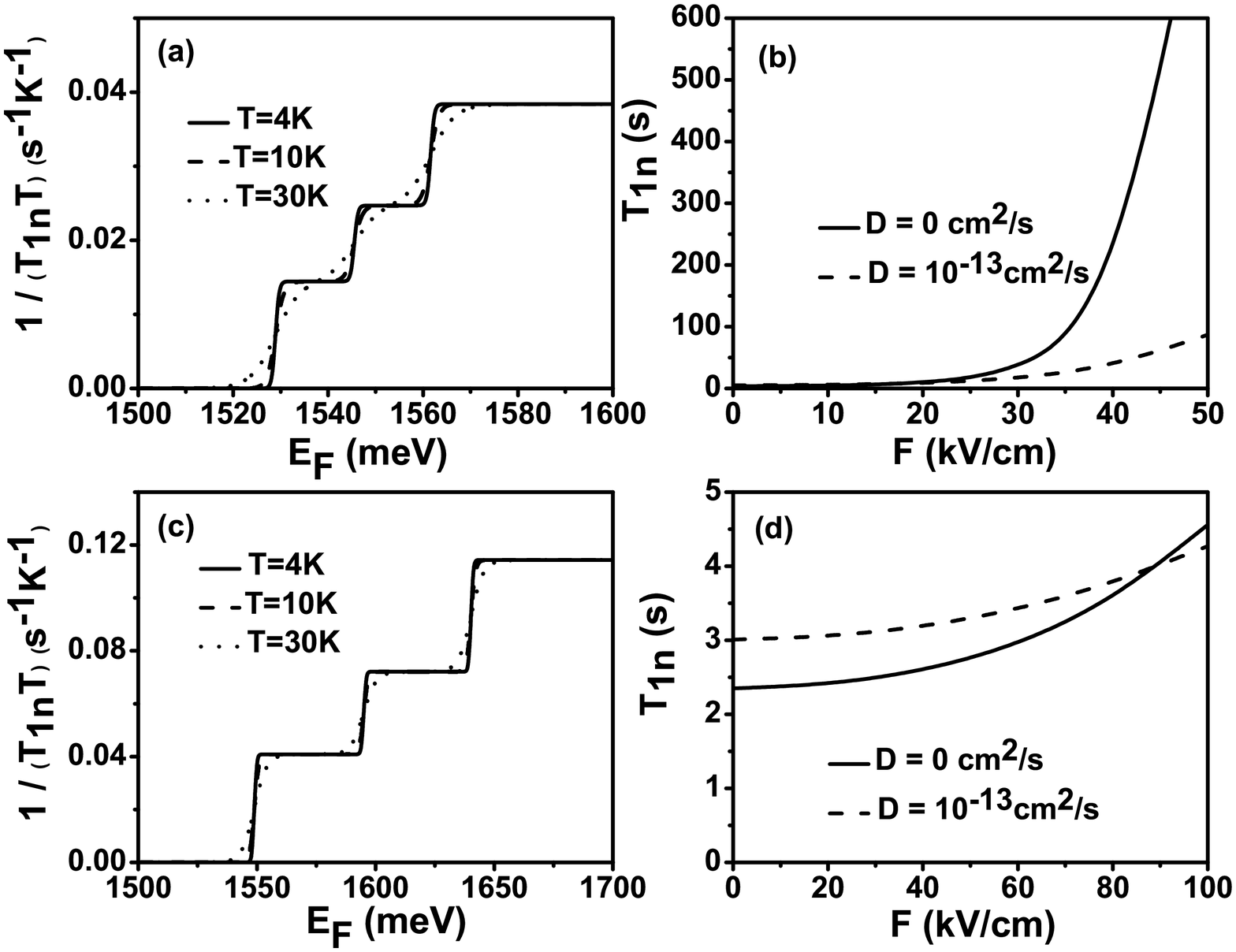}}
%
%
\caption{The \emph{initial} value of $1/(T_{1n}T)$ as a function
of Fermi energy at several temperatures [(a) PQW I and (c) PQW
II]. The electric field dependence of the \emph{initial} nuclear
spin relaxation times for  single subband occupancy, with and
without nuclear spin diffusion [(b) PQW I and (d) PQW II].}
\label{fig3}
\end{figure}

In Fig.~\ref{fig4}(a) and \ref{fig4}(c) we present the position
dependence of $1/(T_{1n}T)$ for different values of the applied
electric field. As seen in the analytic studies, in PQW's the
applied electric field leads to a displacement of the electronic
wave function with direct implications for the position dependent
nuclear spin relaxation time.  Fig.~\ref{fig4}(b) and
\ref{fig4}(d) showcase the field dependence of $1/(T_{1n}T)$ for
three different position across the QW's. These positions were
selected in the same way as those for Fig.~\ref{fig2},
corresponding to maximal and 50\% of maximal initial nuclear
polarization. The electric field tunability is much higher in PQW
I, where the effects of the electric field are considerable even
at low fields. The achieved tunability is a few orders of
magnitude for both structures for any of the three considered
positions.  The results shown in Fig.~\ref{fig4} are very similar
to those obtained with the analytic model and shown in
Fig.~\ref{fig2}. In Fig. \ref{fig3} we present the value of
$1/(T_{1n}T)$ for  uniformly distributed nuclei as a function of
the Fermi energy. Just as for the analytic results presented in
Fig.~\ref{fig1} the temperature-induced smearing is more
significant for PQW I, which has a smaller subband separation,
than for PQW II. However the effects of nuclear diffusion are more
pronounced for PQW I than for PQW II.

\section{Summary}

We have performed a detailed analysis, both analytically and
numerically, of the nuclear spin dynamics as a result of the
hyperfine interaction in parabolic quantum wells. Analytical
formulae for the electron and nuclear spin relaxation times were
obtained for general 2D structures and extended for quasi 2D
structures. The electronic structure enters these formulae in the
form of the electronic local density of states. For the quasi 2D
situation we considered an ideal parabolic QW, for which analytic
results are available for the field-dependent envelope functions.
Thus we were able to explore the electric-field manipulation of
the nuclear spin relaxation time entirely within an analytic
model. These results were checked with  numerical {\bf k$\cdot$p}
calculations for the envelope functions. The results obtained from
analytical and numerical calculations are consistent, with small
differences originating from approximations used in the analytical
calculation.

Our specific calculations were for two different PQW structures,
where the main difference was the value of the characteristic
frequency in the confining potential (or subband energy
separation).  Electrical field control of the nuclear spin
dynamics in PQW structures can be achieved in two different ways:
by controlling the electronic subband occupancy or by controlling
the position dependent electronic wave functions. When the
electronic Fermi energy is changed a stepwise dependence of the
nuclear spin relaxation time is predicted. Our calculation
identifies the PQW with a larger energy splitting between subbands
as more suitable for such manipulation. On the other hand, for
single subband occupancy, a position dependent control of the
nuclear spin relaxation time is possible as the electronic local
density of states across the PQW is tunable by tilting the PQW
potential with an electric field. In this case we suggested that
the insertion of a $\d$-doped layer of different NMR active nuclei
than the host nuclei would allow manipulation of the nuclear spin
relaxation time on the scale of a few orders of magnitude. Here
samples with a small value of the energy splitting between
adjacent electronic subbands are more suitable. The effect of
nuclear spin diffusion on the nuclear spin dynamics for single
subband occupancy was considered for several values of the applied
electric field. We find that the presence of spin diffusion will
diminish the electrical field control of the relaxation process.
However, for the PQW with a smaller energy splitting of the
subbands, electric field tunability of the nuclear spin relaxation
time is more robust.

\begin{acknowledgments}
We would like to acknowledge stimulating discussions with D. D.
Awschalom, W. H. Lau,  D. Loss, A. V. Khaetskii, J. M. Kikkawa and
J. M. Tang. This work was supported by DARPA/ARO DAAD19-01-1-0490.
\end{acknowledgments}

\newpage

\end{document}